\documentclass{moriond}
\usepackage{mathtools}
\DeclareMathAlphabet{\mathcal}{OMS}{cmsy}{m}{n}

\def\Journal#1#2#3#4{{#1} {\bf #2}, #3 (#4)}

\def\PLB{{\em Phys. Lett.}  B}

\def\PRD{{\em Phys. Rev.} D}

\def\ra{\rightarrow}

\def\be{\begin{equation}}
\def\ee{\end{equation}}
\def\bea{\begin{eqnarray}}
\def\eea{\end{eqnarray}}

\newcommand{\pt}{\ensuremath{p_{\mathrm{T}}}}
\newcommand{\ETm}{\ensuremath{E_{\mathrm{T}}^{\text{miss}}}}
\newcommand{\MET}{\ETm}
\newcommand{\cmsSymbolFace}{\mathrm}
\newcommand{\cPqb}{\ensuremath{\cmsSymbolFace{b}}} 
\newcommand{\cPaqb}{\ensuremath{\overline{\cmsSymbolFace{b}}}} 
\newcommand{\bbbar}{\ensuremath{\cPqb\cPaqb}}

\newcommand{\Photo}{\includegraphics[height=35mm]{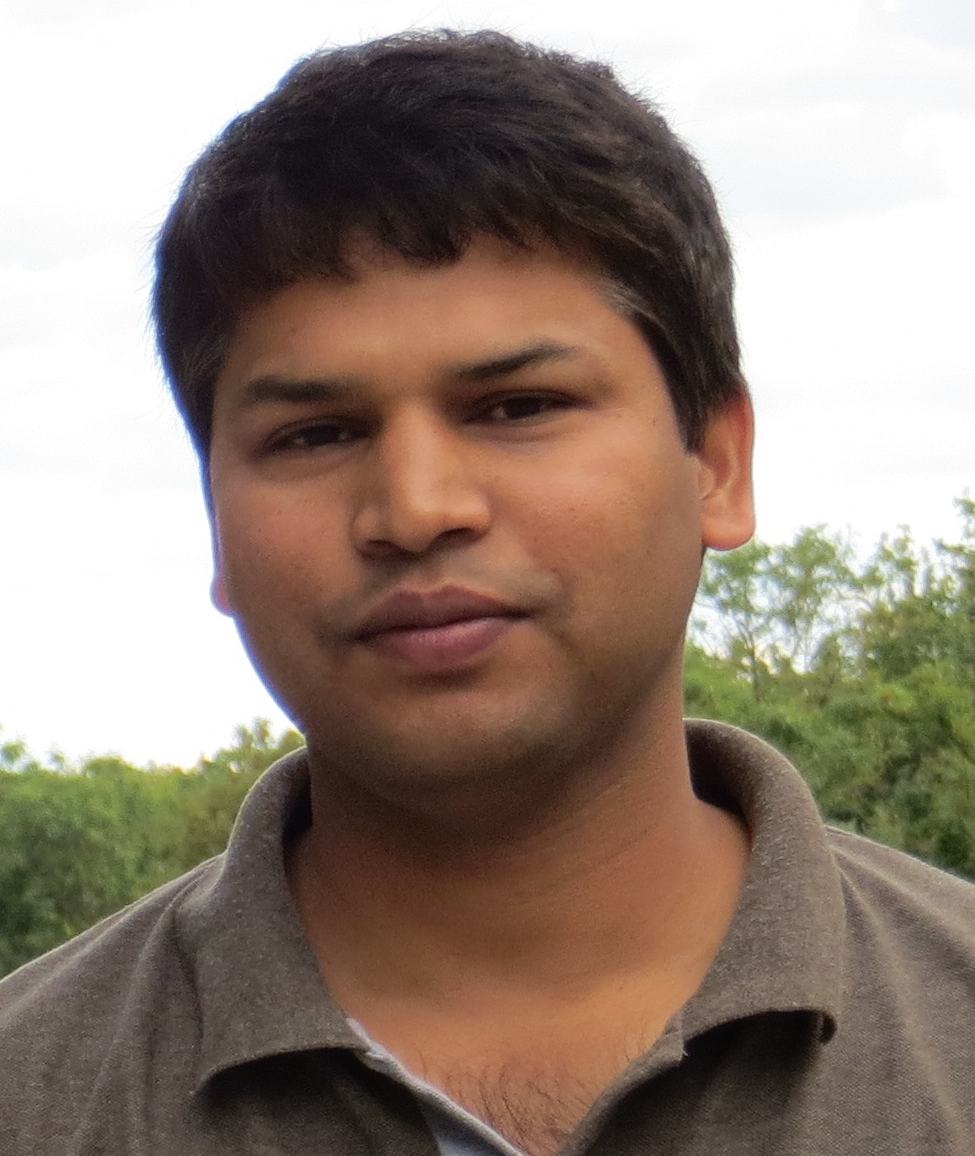}}

\begin{document}
\vspace*{4cm}
\title{Study of Higgs production in fermionic decay channels at the LHC}

\author{ Aruna Kumar Nayak }

\address{Deutsches Elektronen-Synchrotron, Hamburg, Germany \\
 (On behalf of the ATLAS and CMS collaborations)}

\maketitle\abstracts{
  The results of the searches for a Higgs boson decaying to down-type fermions  
  are presented using the proton-proton collision data collected by the ATLAS and CMS experiments at the LHC, 
  at a centre-of-mass energy of 7 \& 8 TeV. The results obtained with the available data provides a strong 
  evidence for a Higgs boson coupling to fermions. Results are also presented for a Higgs boson production in 
  association with a pair of top quarks.  
}

\section{Introduction}
A Higgs boson of mass around 125 GeV has recently been discovered by the ATLAS and CMS experiments at LHC~\cite{atlas_obs,cms_obs}.
The discovery was made in channels where the Higgs boson decays to a pair of gauge bosons. 
However, there was no direct observation in fermionic channels.
Thereafter, the LHC added more data and the analysis strategies for the searches in fermionic final states have been greatly improved. 
In SM, the Higgs boson couples to fermions via Yukawa interactions, where the coupling strength is proportional to the fermion mass.
Thus, ${\rm H}\ra \tau\tau$ and ${\rm H}\ra \bbbar$ are two major fermionic channels to search for the Higgs boson at LHC.

Due to the nature of the Yukawa coupling in the SM, the Higgs boson is expected to have strong coupling to the top quark relative to the other fermions. 
The coupling to the top quark can be accessed indirectly through the measurements of gluon fusion production mechanism and Higgs boson decay to photons, which involves the fermion loop dominated by the top quark contribution.
The current measurements of top quark Yukawa couplings are consistent with that of SM predictions within the experimental uncertainties~\cite{atlas_comb,cms_comb}. 
However, the production of the Higgs boson in association with a top-quark pair ($t\bar{t}{\rm H}$) allows for a direct probe of top-quark Yukawa coupling. 
A measurement of the $t\bar{t}{\rm H}$ production rate will provide a direct test of the Higgs boson's coupling to top quark. 

This article summarizes the results of the searches for the SM Higgs boson in the channels ${\rm H}\ra \bbbar$, ${\rm H}\ra \tau\tau$, ${\rm H}\ra \mu\mu$, and ${\rm H}\ra ee$ with the data collected by the ATLAS and CMS experiments at LHC at a centre-of-mass energy of 7 \& 8 TeV, corresponding to an integrated luminosity of approximately 5 \& 20 fb$^{-1}$, respectively.
The results of a search for $t\bar{t}{\rm H}$ production in the above experiments are also described.

\section{Searches for $V{\rm H},~{\rm H}\ra \bbbar$}\label{sec:hbb}
Since the inclusive production of ${\rm H}\ra \bbbar$ is overwhelmingly dominated by the QCD multi-jets background, the search is performed in the production mode where the Higgs boson is produced in association with a vector boson~\cite{atlas_hbb,cms_hbb}. 
The following search channels are considered: W($\mu\nu$)H, W(e$\nu$)H, W($\tau\nu$)H, Z($\mu\mu$)H, Z(ee)H, and Z($\nu\nu$)H, where H decays to $\bbbar$. 
The major backgrounds arise from the production of W/Z+jets, $t\bar{t}$, dibosons and QCD multi-jet events. 
The analysis strategy is based on the reconstruction of the vector bosons in their leptonic decay modes and of the Higgs boson decay into two $\cPqb$-tagged jets. 
The events are further categorized based on the $\pt$ of the vector bosons, and $\cPqb$-tagging discriminator values of the jets. 
The ${\rm H}\ra \bbbar$ decay is reconstructed by selecting a pair of $\cPqb$-tagged jets, for which $\pt$(jj) is highest. 
Both the experiments have developed techniques to improve the $\cPqb$-jet energy calibration, thus improving the Higgs boson mass resolution.
The sensitivity of the search is improved using a multivariate analysis (MVA) approach.
Dedicated Boosted Decision Tree (BDT) discriminants are trained combining dijet mass with other variables sensitive to kinematic, topological and $\cPqb$-tagging properties of the selected events.

The CL$_{s}$ method~\cite{cls_method} is used to obtain 95\% confidence level (CL) upper limits on $\sigma / \sigma_{SM}$. 
For a Higgs boson of mass 125 GeV, the observed (expected) limits are 1.2 (0.8) and 1.89 (0.95), obtained by ATLAS and CMS experiments, respectively.
The best fit values of the signal strength parameter, $\mu$ ($\sigma / \sigma_{SM}$), for $m_{{\rm H}}$ = 125 GeV are found to be $0.51\pm0.31{\rm (stat.)}\pm0.24{\rm (syst.)}$ by the ATLAS, and $0.84\pm0.44$ by the CMS experiment, respectively. 
The ATLAS measured an observed local significance of 1.4$\sigma$, to be compared to an expectation of 2.6$\sigma$, for a Higgs boson mass of 125 GeV, while the CMS found an observed (expected) significance of 2.0$\sigma$ (2.6$\sigma$). 
Note that the values of $\mu$ and significance obtained by the CMS also includes the production process $t\bar{t}{\rm H},~{\rm H}\ra \bbbar$ and $gg\ra {\rm Z}{\rm H},~{\rm H}\ra \bbbar$~\cite{cms_comb}, while the later production process is included in the ATLAS result.
All these results are compatible with the presence of the SM Higgs boson of mass approximately 125 GeV. 

\section{Searches for ${\rm H}\ra \tau^{+}\tau^{-}$}\label{sec:htautau}
Both the ATLAS and CMS experiments have searched for the SM Higgs boson decaying to a pair of $\tau$ leptons~\cite{atlas_htt,cms_htt}.
The search strategy for ${\rm H}\ra \tau\tau$ makes use of all leading production mechanisms at LHC: gluon-gluon fusion (ggF), vector boson fusion (VBF), and associated production with a W or Z boson. 
The analyses consider all possible decays of $\tau$ lepton, hadronically or leptonically, leading to six different final states: $\mu\tau_{h}$, e$\tau_{h}$, $\tau_{h}\tau_{h}$, e$\mu$, $\mu\mu$, and ee, where $\tau_{h}$ denotes the hadronic decay of $\tau$ lepton. 
The major backgrounds arise from the production of ${\rm Z}\ra \tau\tau$, W+jets, $t\bar{t}$+jets and QCD multi-jet events. 
Most of the background yields are estimated, and their distributions are modeled, using data.

To maximize the sensitivity of the analysis, the ATLAS experiment employs a MVA approach. 
The events are divided in to two broad categories: VBF and boost category. 
The VBF category targets events produced in the VBF production mechanism, while the boost category targets events with a boosted Higgs boson produced in ggF process. 
Separate BDTs are trained for each analysis channel and category, for a Higgs boson mass of 125 GeV, combining the variables sensitive to the kinematic properties of the leptons and jets, $\MET$, and the reconstructed invariant mass of the $\tau$-lepton pair ($m_{\tau\tau}$).  
In both the experiments, $m_{\tau\tau}$ is computed using likelihood based techniques from the four-momenta of the individual leptons and the $\MET$ vector. 
The CMS experiment adopted a cut-based analysis approach to enhance the sensitivity of the analysis. 
The events are divided in to several categories using the variables such as number of jets, dijet mass ($m_{jj}$), $\eta$ difference between the jets ($\Delta\eta_{jj}$), visible $\pt$ of the $\tau$ leptons, and $\pt$ of the reconstructed di-$\tau$ candidates.
The distribution of the BDT discriminant and $m_{\tau\tau}$ are used as the final discriminator in ATLAS and CMS analyses, respectively. 

\begin{figure}
\begin{minipage}{0.5\linewidth}
\centerline{\includegraphics[width=0.65\linewidth]{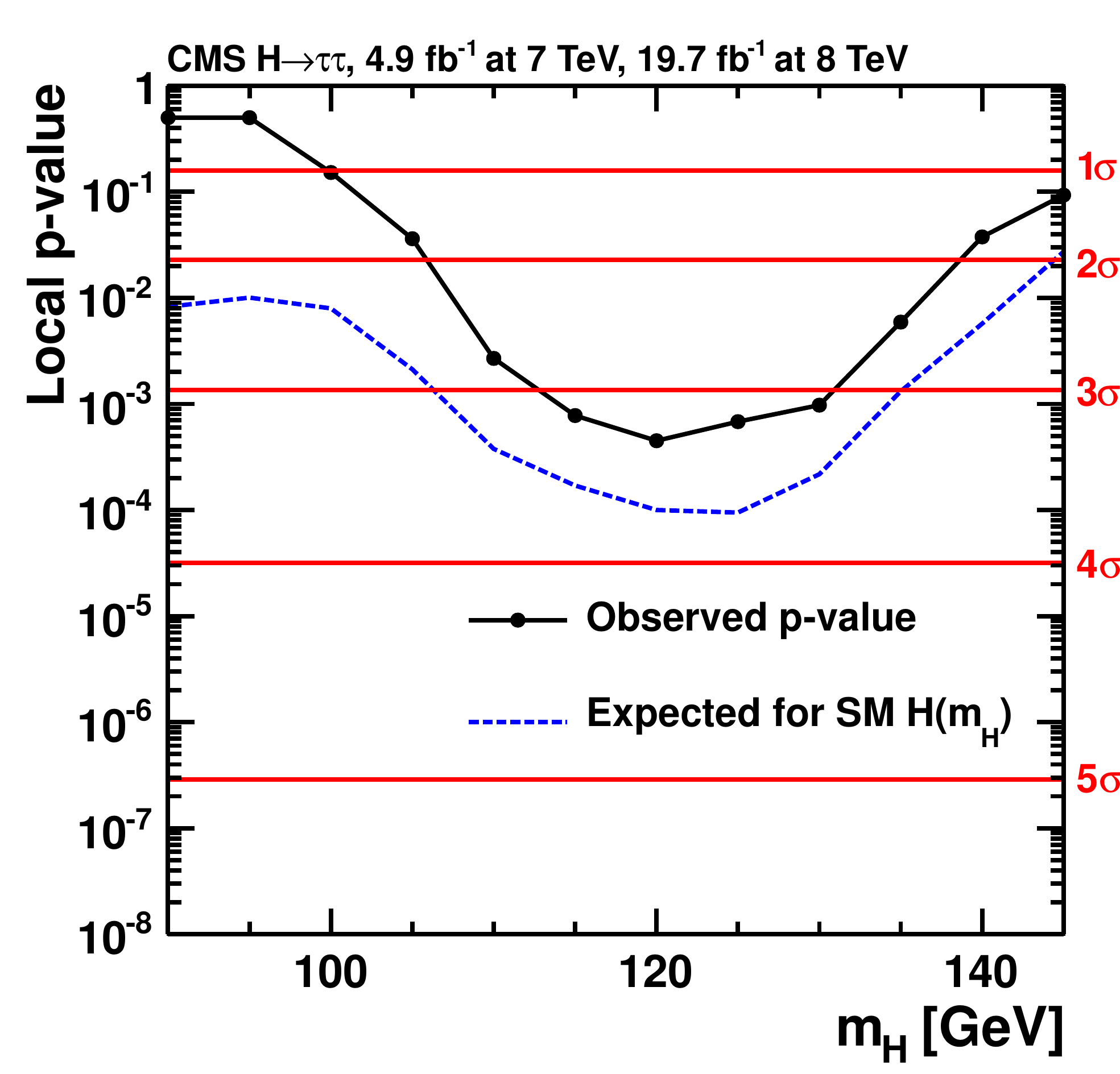}}
\end{minipage}
\hfill
\begin{minipage}{0.5\linewidth}
\centerline{\includegraphics[width=0.65\linewidth]{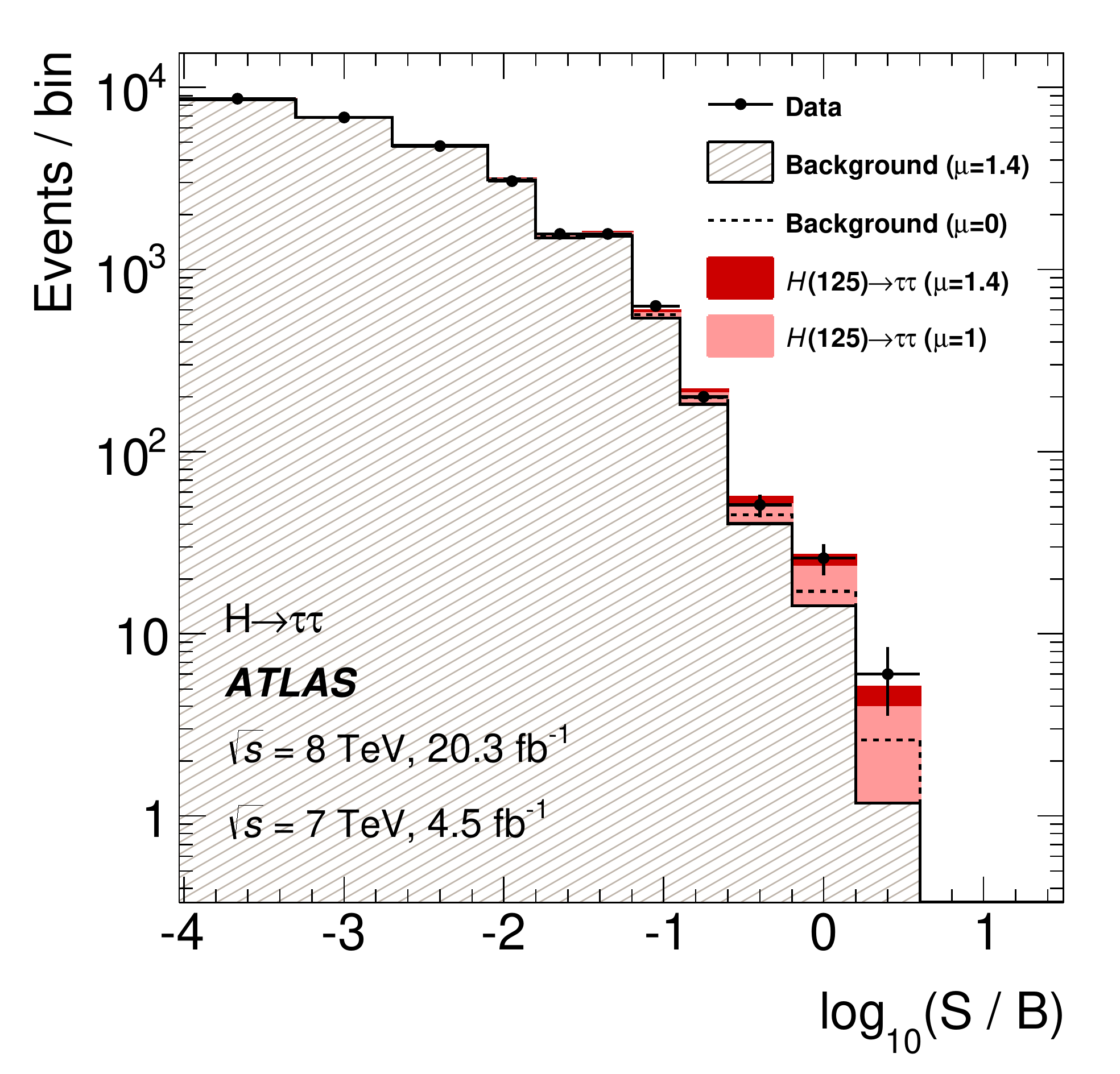}}
\end{minipage}
\caption[]{Left: Local $p$-value and significance as function of SM Higgs boson mass hypothesis from CMS analysis. Right: Event yields as a function of log$_{10}$(S/B), where S (signal yield) and B (background yield) are taken from BDT output, assuming $\mu$ = 1.4.}
\label{fig:htautau}
\end{figure}

An excess of events is observed with respect to the background-only hypothesis, which is compatible with the presence of the SM Higgs boson of mass 125 GeV.
The excess is quantified by the local p-values, which is shown in Fig.~\ref{fig:htautau}(left) for CMS analysis.
For $m_{{\rm H}}$ = 125 GeV, the expected and observed p-values correspond to a significance of 3.7 and 3.2 standard deviations, respectively. 
The corresponding values from ATLAS analysis are 3.4$\sigma$ (expected) and 4.5$\sigma$ (observed) at $m_{{\rm H}}$ = 125.36 GeV. 
The best fit values of $\mu$ are 1.43$^{+0.43}_{-0.37}$ at $m_{{\rm H}}$ =125.36 GeV and 0.78 $\pm$ 0.27 at $m_{{\rm H}}$ = 125 GeV from ATLAS and CMS analyses, respectively. 
The distribution of event yield, in bins of log$_{10}$(S/B), for all signal region bins is shown in Fig.~\ref{fig:htautau}(right) for ATLAS analysis, which is calculated assuming $\mu$ = 1.4.
These observations provide a strong evidence for the coupling of the Higgs boson to $\tau$ leptons as expected from the SM. 

The CMS experiment combined the results of the searches for a Higgs boson decaying to $\cPqb$ quarks and $\tau$ leptons~\cite{cms_fermion}, which results in a strong evidence for the direct coupling of the observed 125 GeV Higgs boson to down-type fermions. 
The observed significance for a Higgs boson to down-type fermion coupling is 3.8$\sigma$, to be compared to an expectation of 4.4$\sigma$. 

\section{Searches for ${\rm H}\ra \mu^{+}\mu^{-}$ \& ${\rm H}\ra {\rm e}^{+}{\rm e}^{-}$ }
The CMS experiment performed a search for a SM-like Higgs boson decaying to a pair of muons or a pair of electrons~\cite{cms_hmm}, while the ATLAS experiment searched for a Higgs boson decaying to a pair of muons~\cite{atlas_hmm}. 
No excess of events are observed, and a 95\% CL upper limit is obtained on the production cross section times branching fraction using the CL$_{s}$ method.
For a SM Higgs boson with a mass of 125 GeV (125.5 GeV for ATLAS), the upper limit obtained on branching fractions are $\mathcal{B}({\rm H}\ra \mu^{+}\mu^{-})<$ 0.0016 and $\mathcal{B}({\rm H}\ra {\rm e}^{+}{\rm e}^{-})<$ 0.0019 from CMS analysis, and $\mathcal{B}({\rm H}\ra \mu^{+}\mu^{-})<$ 0.0015 from ATLAS analysis, respectively. 
These results provide a confirmation for the fact that the Higgs boson coupling to leptons are not flavor universal, unlike vector bosons. 

\section{Searches for $t\bar{t}{\rm H}$ production}
The search for a Higgs boson produced in association with a pair of top quarks at LHC is extremely challenging due to its small production cross section, approximately 130 fb at 8 TeV, and large background contribution from inclusive $t\bar{t}$+jets production process.
The searches are performed in final states where W bosons from both the top quarks decay to leptons, or W boson from one top quark decays to lepton while the other decays to jets.
The ATLAS experiment performed searches for $t\bar{t}{\rm H}$ production where the Higgs boson decays to $\bbbar$~\cite{atlas_tthbb}, and where the Higgs boson decays to leptons via the decay of ${\rm W}{\rm W}^{*}$, $\tau\tau$, and ${\rm Z}{\rm Z}^{*}$~\cite{atlas_tthlep}. 
The CMS experiment performed searches for $t\bar{t}{\rm H}$ production where the following signatures of Higgs boson decay are considered: ${\rm H}\ra$ hadrons, ${\rm H}\ra$ photons, and ${\rm H}\ra$ leptons~\cite{cms_tth}. 
These decays proceed via the decay of Higgs boson to $\bbbar$, $\tau\tau$, ${\rm W}{\rm W}^{*}$, ${\rm Z}{\rm Z}^{*}$, and $\gamma\gamma$. 

The general strategy is to select $t\bar{t}$-like events requiring at least one or more $\cPqb$-tagged jets. 
To enhance the signal over $t\bar{t}$ background, the events are divided in to several categories based on the number of jets and $\cPqb$-tagged jets. 
Both experiments use MVA approach to separate signal from backgrounds. 

\begin{figure}
\begin{minipage}{0.32\linewidth}
\centerline{\includegraphics[width=0.8\linewidth]{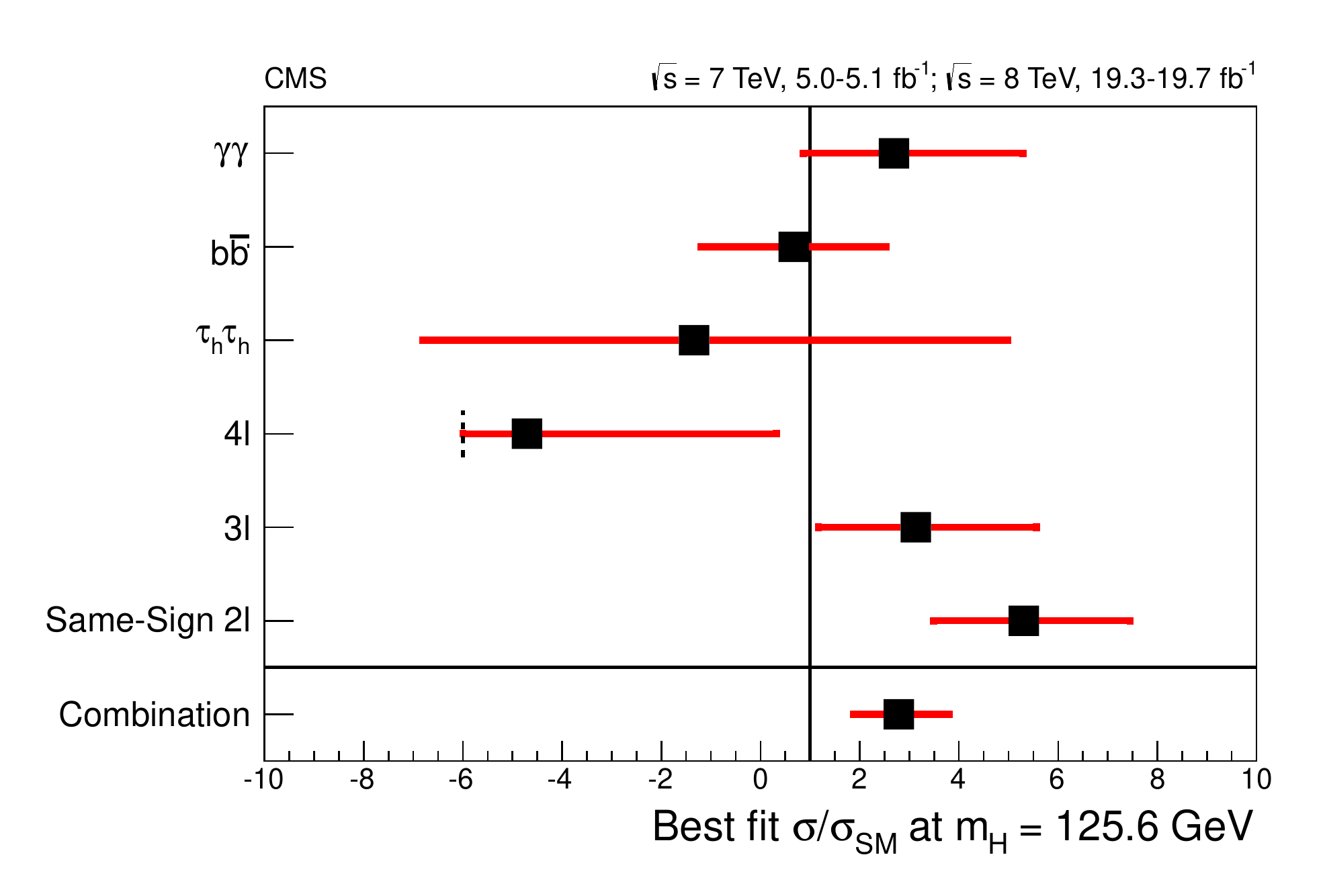}}
\end{minipage}
\hfill
\begin{minipage}{0.32\linewidth}
\centerline{\includegraphics[width=0.8\linewidth]{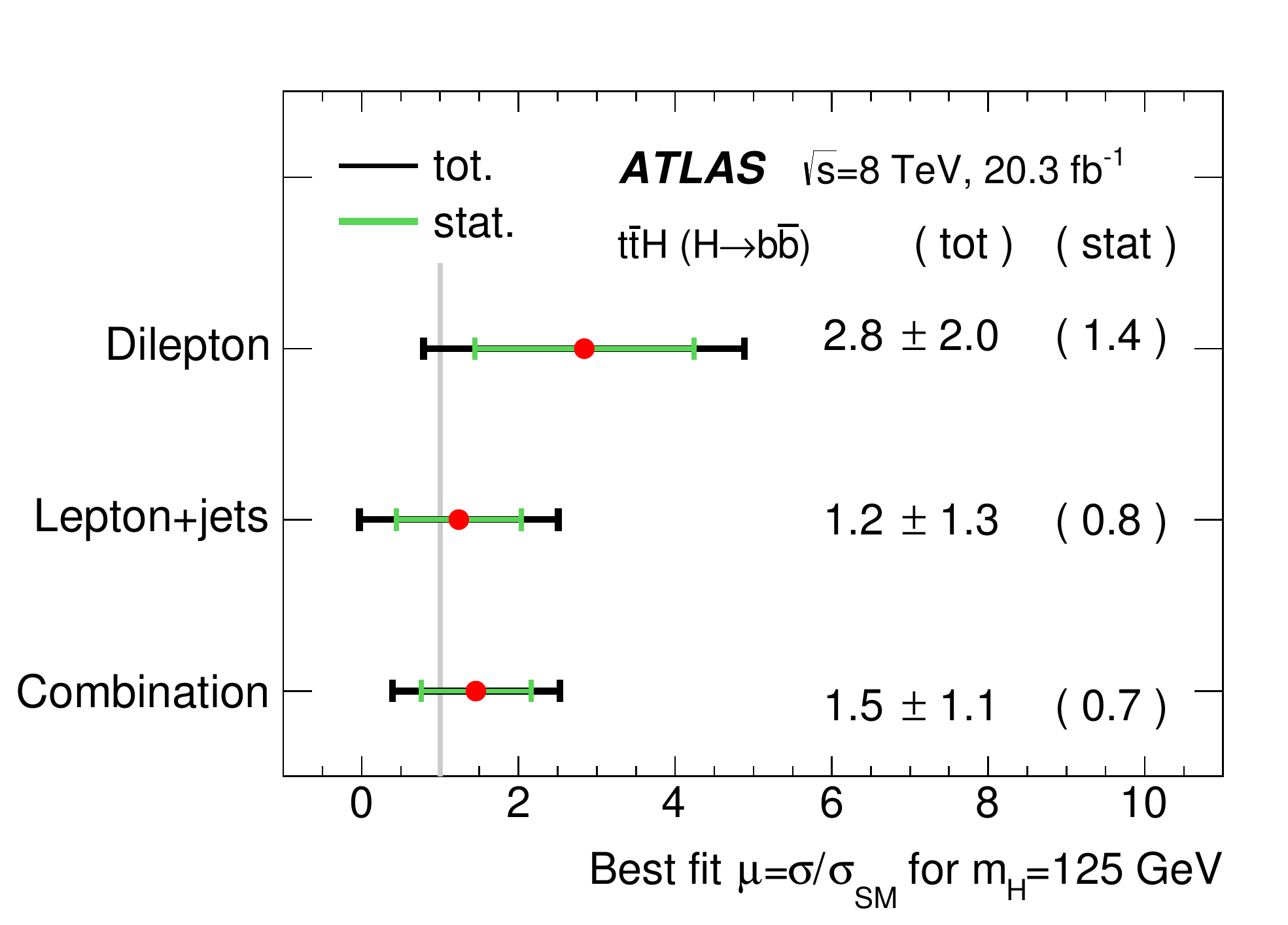}}
\end{minipage}
\hfill
\begin{minipage}{0.32\linewidth}
\centerline{\includegraphics[width=0.8\linewidth]{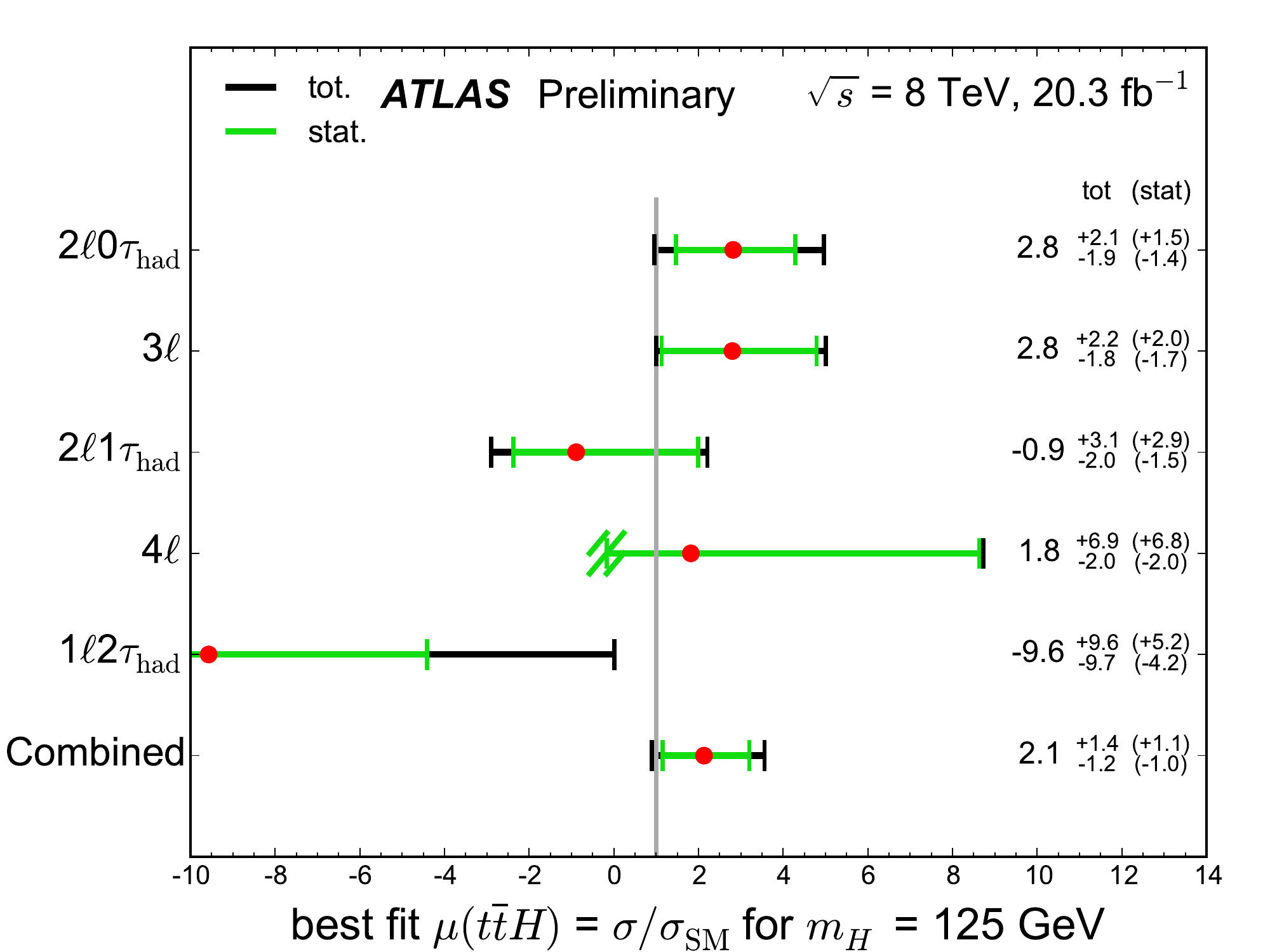}}
\end{minipage}
\caption[]{The best fit value of the signal strength parameter $\mu = \sigma / \sigma_{SM}$ for each $t\bar{t}{\rm H}$ channel.}
\label{fig:tth}
\end{figure}

The combined CMS results obtained an observed (expected) upper limit of 4.5 (1.7) on $\sigma / \sigma_{SM}$ at 95\% CL.
The upper limits obtained by ATLAS are compatible with the presence of a SM Higgs boson. 
Figure~\ref{fig:tth} shows measured values of signal strength $\mu$ in both the experiments. 
While the measured $\mu$ values in ATLAS are consistent with SM expectations within the uncertainty, the combined value of $\mu$ obtained in CMS have a roughly 2 standard deviation upward fluctuation. 
The CMS experiment also performed a separate analysis for the search of $t\bar{t}{\rm H},~{\rm H}\ra \bbbar$ using matrix-element method~\cite{cms_tth_matrix}, which improves the sensitivity by approximately 15\% compared to the standard $t\bar{t}{\rm H},~{\rm H}\ra \bbbar$ analysis. 
The ATLAS experiment has also utilized matrix-element discriminators in some of the high purity categories of the $t\bar{t}{\rm H},~{\rm H}\ra \bbbar$ analysis, which were used as input to the multivariate discriminant. 

\section{Conclusion}
The results of the searches for the SM Higgs boson decaying to a pair of $\cPqb$ quarks or a pair of $\tau$ leptons using the ATLAS and CMS detectors provide a strong evidence for the Higgs boson coupling to down-type fermions. 
The experiments have made strong progress in the direct measurement of Higgs boson coupling to top quark. 
The results obtained with the available data are consistent with the SM expectations. 
The LHC run at 13 TeV, expected to start from mid 2015, will provide definitive answers about the Higgs boson coupling to fermions.


\section*{References}

\begin{thebibliography}{99}
\bibitem{atlas_obs}ATLAS Collaboration, \Journal{\PLB}{716}{1}{2012}.

\bibitem{cms_obs}CMS Collaboration, \Journal{\PLB}{716}{30}{2012}.

\bibitem{atlas_comb}ATLAS Collaboration, ATLAS-CONF-2015-007 (2015).

\bibitem{cms_comb}CMS Collaboration, arXiv:1412.8662 [hep-ex] (2014).

\bibitem{atlas_hbb}ATLAS Collaboration, \Journal{\em JHEP}{01}{069}{2015}.

\bibitem{cms_hbb}CMS Collaboration, \Journal{\PRD}{89}{012003}{2014}.

\bibitem{cls_method}A. L. Read, \Journal{\em J. Phys. G}{28}{2693}{2002}.

\bibitem{atlas_htt}ATLAS Collaboration, \Journal{\em JHEP}{04}{117}{2015}.

\bibitem{cms_htt}CMS Collaboration, \Journal{\em JHEP}{05}{104}{2014}.

\bibitem{cms_fermion}CMS Collaboration, \Journal{\em Nature Physics}{10}{557}{2014}.

\bibitem{cms_hmm}CMS Collaboration, \Journal{\PLB}{744}{184}{2015}.

\bibitem{atlas_hmm}ATLAS Collaboration, \Journal{\PLB}{738}{68}{2014}.

\bibitem{atlas_tthbb}ATLAS Collaboration, arXiv:1503.05066 [hep-ex] (2015).

\bibitem{atlas_tthlep}ATLAS Collaboration, ATLAS-CONF-2015-006 (2015).

\bibitem{cms_tth}CMS Collaboration, \Journal{\em JHEP}{09}{087}{2014}.

\bibitem{cms_tth_matrix}CMS Collaboration, arXiv:1502.02485 [hep-ex] (2015).

\end{thebibliography}
\end{document}